\documentclass{emulateapj}

\received{2011 July 8}
\accepted{2011 August 15}
\slugcomment{To appear in {\it the Astrophysical Journal, Letter}}

\shortauthors{KOBAYASHI et al.}
\shorttitle{Evolution of Fluorine Abundances with the $\nu$-Process}

\def\gtsim {>\kern-1.2em\lower1.1ex\hbox{$\sim$}~}   
\def\ltsim {<\kern-1.2em\lower1.1ex\hbox{$\sim$}~}   

\begin{document}

\def \pasa {PASA}

\title{Evolution of Fluorine in the Galaxy with the $\nu$-Process}
\author{Chiaki KOBAYASHI$^1$, Natsuko IZUTANI$^2$, Amanda I. KARAKAS$^1$, Takashi YOSHIDA$^2$, David YONG$^1$, and Hideyuki UMEDA$^2$}
\affil{$^1$ Research School of Astronomy \& Astrophysics, The Australian National University, Cotter Rd., Weston CT 2611, Australia; chiaki@mso.anu.edu.au}
\affil{$^2$ Department of Astronomy, School of Science,
University of Tokyo, Bunkyo-ku, Tokyo 113-0033, Japan}

\begin{abstract}
We calculate the evolution of fluorine in the solar neighborhood with the $\nu$-process of core-collapse supernovae, the results of which are in good agreement with the observations of field stars.
The $\nu$-process operating in supernovae causes the [F/O] ratio to plateau at [O/H] $\ltsim -1.2$, followed by a rapid increase toward [O/H] $\sim -0.5$ from the contribution of Asymptotic Giant Branch stars.
The plateau value of [F/O] depends on the neutrino luminosity released by core-collapse supernovae and may be constrained by using future observations of field stars at low metallicities.
For globular clusters, the handful of [F/O] measurements suggest that the relative contribution from low-mass supernovae is smaller in these systems than in the field.
\end{abstract}


\section{Introduction}

Most of the energy from core-collapse supernovae is released as neutrinos and anti-neutrinos ($\gtsim 10^{53}$ erg). However, the interaction of the neutrinos with matter and the effects on the nucleosynthesis have only been discussed for a few models \citep[e.g.,][]{woo90,woo95,yos04,heg05,yos08,nak10}.
The $\nu$-process does not affect the yields of major elements such as Fe and $\alpha$ elements, but it will increase those of some elements such as B, F, K, Sc, V, Mn, and Ti.

Fluorine is an intriguing, though currently poorly studied element. Most studies of F are from cool stars in which F measurements are only available from the HF molecule near 2.3 microns. Often, the F abundance comes from a single HF line.
Asymptotic Giant Branch (AGB) stars and massive stars have both been suggested to produce F \citep{jor92} but the F production has only been confirmed for AGB stars \citep[and references therein]{wer05,wer09,abia10,abia11,otsuka11}.
Both in low-mass and massive stars, $^{19}$F is produced by core and shell He-burning at $T\gtsim$ $1.5 \times 10^8$ K, but is destroyed by $\alpha$-captures once the temperatures exceed $\sim 2.5 \times 10^{8}$ K.
In AGB stars, there is a primary component produced by the $^{18}$O(n, $\gamma$)$^{19}$O($\beta^-$)$^{19}$F reaction \citep{gal10}, which is included in the network used to compute the AGB yields \citep{kar10}.
AGB models with initial masses of $\sim 4-7 M_\odot$ destroy F by proton captures that occur at the base of the convective envelope (hot bottom burning).
In AGB stars the production of F is highly mass dependent, where F production peaks at $\sim 3 M_\odot$ at solar metallicity \citep{lugaro04}.

\citet{kob11} showed that since the AGB mass range that produces F is $2-4 M_\odot$, this contribution is seen only at [Fe/H] $\gtsim -1.5$ in Galactic chemical evolution models \citep[see also][]{tra99}, and that the F production from AGB stars is not enough to explain the observations around [Fe/H] $\sim 0$.
In the other Galactic chemical evolution models, \citet{tim95} showed that their F yields of core-collapse supernovae with the $\nu$-process were not enough to meet the observations of stars.
Massive stars evolving as Wolf-Rayet (WR) stars will experience very strong stellar winds, which may prevent the destruction of F \citep{mey00}.
\citet{ren04} showed that [F/O] could be enhanced at [O/H] $\gtsim -0.2$ with the WR yields of \citet{mey00} in addition to the AGB yields.
However, the contribution by WR to F may be reduced by including rotation in the stellar models \citep{heg05b,pal05}.
Therefore, we do not include the yields of WR stars in this paper.

In this paper we show the effects of the $\nu$-process in the Galactic chemical evolution of fluorine using latest yields of core-collapse supernovae and AGB stars. The F yields of SNe Ia are also included but are very small.
In \S 2 we briefly describe our supernova models, where the details of the models will be described in Izutani et al. in preparation.
In \S 3 we show the results of our chemical evolution models of the solar neighborhood including our nucleosynthesis yields with the $\nu$-process. 
\S 4 denotes our conclusions and discussion.

\section{The $\nu$-process}

Although the cross sections of neutrino-nucleus reactions are small, a large flux of nutrinos is released when the core of a massive star collapses to form a neutron star.
For this reason, the $\nu$-process can have a significant effect on the nucleosynthesis of core-collapse supernovae.
Neutrinos are emitted not only from a collapsing proto-neutron star but also from the innermost region just above a black hole \citep{sur05}.
We adopt the $\nu$-process up to $^{80}$Kr in our nucleosynthesis calculations as in \citet{yos08}, both for the cases of supernovae (SNe, the explosion energy of $E = 10^{51}$ erg) and hypernovae (HNe, $E > 10^{51}$ erg).
The neutrino luminosity is assumed to be uniformly partitioned among the neutrino flavors, and is assumed to decrease exponentially in time with a timescale of 3 sec \citep{woo90}.
The total neutrino energy is given by a free parameter and in this paper we present two cases with $E_\nu = 3 \times 10^{53}$ erg, which corresponds to the gravitational binding energy of a $1.4M_\odot$ neutron star \citep{lat01}, and $9 \times 10^{53}$ erg as the maximum possible effect of the $\nu$-process.
The neutrino energy spectra are assumed to be Fermi-Dirac distributions with zero chemical potentials.
The temperatures of $\nu_{\mu,\tau}$, $\bar{\nu}_{\mu,\tau}$ and $\nu_e$, $\bar{\nu}_e$ are set to be $T_\nu = 6$ MeV$/k$ and $4$ MeV$/k$, respectively \citep{rau02}.
Note that the $\nu$-cross sections contain some uncertainties \citep{heg05}.

In a supernova, neutrinos interact with heavy elements through 
neutral-current reactions,
and scatter off nuclei in or near their ground state,
which lead to the excitation of particle unbound states
that decay by neutron, proton, or $\alpha$ emission:
\begin{eqnarray}
(Z,A)+\nu \rightarrow (Z,A)^* + \nu' & \rightarrow & (Z,A-1) + n + \nu' \\
 & \rightarrow & (Z-1,A-1) + p + \nu' \\
 & \rightarrow & (Z-2,A-4) + \alpha + \nu'
\end{eqnarray}
Charged-current reactions of $\nu_e$ or $\bar{\nu}_e$
with heavy nuclei also play a role in producing new elements.
These reactions correspond to the inverse processes of electron
or positron captures.
The new products in excited states emit $\gamma$-rays, neutron, 
proton, or $\alpha$ particles to decay to the ground state.
The capture reactions of the protons and neutrons produced
though these neutrino reactions also enhance the abundances of 
some elements.
For most nuclei, neutral-current reactions are dominant because of the contribution from all flavors of neutrinos and higher temperature of $\nu_{\mu,\tau}$ and $\bar{\nu}_{\mu,\tau}$ than that of $\nu_{e}$ and $\bar{\nu}_e$.

We calculate the nucleosynthesis of core-collapse supernovae with progenitor masses of $M=15,25$, and $50M_\odot$ and initial metallicities of $Z=0,0.004$, and $0.02$ for SNe and HNe.
The nuclear network includes 809 species up to $^{121}$Pd \citep{izu09,izu10}.
The yields are calculated with the same assumptions as in \citet{kob06}:
for SNe, the mass-cut is set to meet the observed iron mass of $0.07M_\odot$.
For HNe, the explosion energy is set to be $10 \times 10^{51}$ and $40 \times 10^{51}$ erg for 25 and $50M_\odot$, respectively, and the parameters of mixing fallback models are determined to get [O/Fe] $=0.5$.
Although there may be diversity in the mixing-fallback process (as in the case of faint supernovae, e.g., \citealt{kob11a}), in this paper we focus on ``typical'' supernovae that are dominant in the Galactic chemical evolution.

In massive stars $^{19}$F is mainly produced in a convective He shell as a secondary product through $^{15}$N$(\alpha,\gamma)^{19}$F, where the F yields are highly dependent on the metallicity.
With the $\nu$-process $^{19}$F is produced in the O- and Ne-enriched region through $^{20}$Ne$(\nu,\nu'p)^{19}$F, and the F yield is increased by a factor of $\sim 10$ and $1000$ for $Z=0.02$ and $Z=0$, respectively.
In the yields, the F/O ratio is smaller for more massive progenitors because of the larger mantle mass and larger O production,
although the mass dependence of F/Fe is not so large.
The F/O ratio does not strongly depend on the explosion energy, but F/Fe is smaller for HNe than SNe II because of the larger Fe production of HNe.

\section{Galactic chemical evolution}

We adopt the $\nu$-process nucleosynthesis yields in the Galactic chemical evolution models.
The nucleosynthesis yields of AGB stars ($1-7M_\odot$) from \citet{kar10} are also included.
We adopt the Kroupa initial mass function (IMF) and the same infall and star formation history as in \citet{kob11}, which reproduces the observed metallicity distribution function (MDF) in the solar neighborhood. 

\begin{figure}
\includegraphics[width=8.5cm]{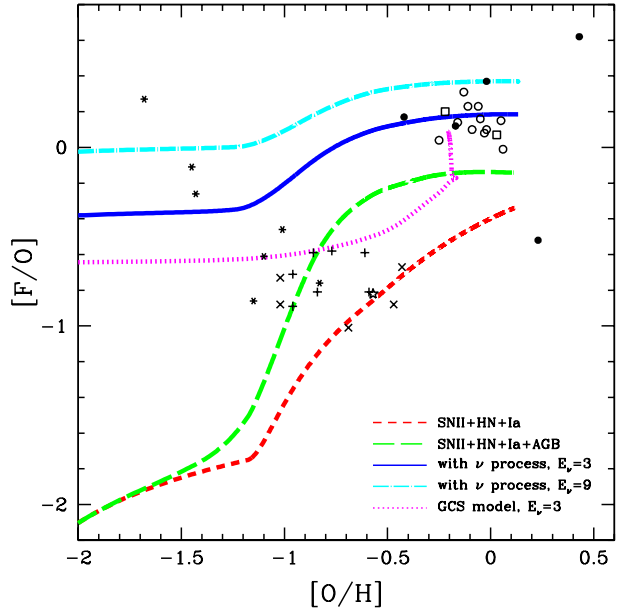}
\caption{\label{fig:fo}
Evolution of the [F/O] ratio against [O/H]
for the solar neighbourhood
with SNe II, HNe, and SNe Ia only (short-dashed lines), 
with AGB stars (long-dashed lines),
with the $\nu$-process of SNe II and HNe (solid line and dot-dashed line for $E_\nu=3 \times 10^{53}$ and $9 \times 10^{53}$ erg, respectively).
The dotted line is for the model for globular clusters.
The observational data sources are:
open circles, \citet{cun03},
open squares, \citet{cun05} for the solar neighbourhood stars;
filled circles, \citet{cun08} for bulge stars.
For the stars in globular clusters,
crosses, \citet{yon08}, NGC 6712;
plus, \citet{smi05}, M4;
stars, \citet{cun03}, $\omega$ Cen;
asterisks, Alves-Brito et al. (2011, in preparation), M22.
}
\end{figure}

Figure \ref{fig:fo} shows the evolution of [F/O] against [O/H].
Without the AGB yields and the $\nu$-process (short-dashed line), the predicted F abundance is too low to meet the observational data at all metallicities.
With the AGB yields (long-dashed line), [F/O] shows a rapid increase from [O/H] $\gtsim -1.2$ toward higher metallicities, which corresponds to the timescale of $2-4 M_\odot$ stars in the solar neighborhood.
At [O/H] $\sim 0$, [F/O] reaches $-0.14$, which is $0.26$ dex larger than the case without the AGB yields.
However, the present [F/O] ratio is still significantly lower than the observations at [O/H] $\sim 0$.
Note that compared to the yields from \citet{kar07},
the F yields from AGB stars in \citet{kar10} were increased 
by applying the slower $^{19}{\rm F}(\alpha,p)^{22}{\rm Ne}$ reaction rate \citep{uga08}.
AGB stars may have polluted some Carbon-Enhanced Metal Poor (CEMP) stars with F at low metallicity via binary interactions \citep{lugaro08,luc11}, or through inhomogeneous enrichment.
However, the overall contribution from AGB stars to the chemical evolution of the Galaxy is minimal at [Fe/H] $\gtrsim -1.5$.

The timescale of supernovae is much shorter than AGB stars, which means that the [F/O] ratio at low metallicities can be strongly enhanced by the $\nu$-process occurring in core-collapse supernovae.
With the standard case of $E_\nu=3 \times 10^{53}$ erg (solid line), the [F/O] ratio shows a plateau of [F/O] $\sim -0.4$ at [O/H] $\ltsim -1.2$, and reaches [F/O] $\sim +0.19$ at [O/H] $\gtsim 0$.
This is consistent with the observational data of field stars at $-0.5 \ltsim$ [O/H] $\ltsim 0$ \citep{cun03,cun05,cun08}.
If we adopt a larger neutrino luminosity of $E_\nu=9 \times 10^{53}$ erg (dot-dashed line), [F/O] can be as large as $\sim +0.37$ at [O/H] $\sim 0$.

In the bulge the star formation timescale is shorter and the average metallicity is higher than the solar neighborhood, but the [F/O] ratio is not so different at [O/H] $\sim 0$ (see Fig. 16 in \citealt{kob11}).
The observations for the bulge stars (filled circles) might suggest that the IMF is also different, although the number of observations is too small to make a conclusion.

At $-1 \ltsim$ [F/O] $\ltsim -0.5$ the observational data are for stars in globular clusters (GCs), where the star formation and chemical enrichment histories are likely to be different to the solar neighborhood.
These GC data seem to be more consistent with the models with the AGB yields only than with the $\nu$-process.
However, it is unlikely that the existence of the $\nu$-process depends on the environment.
With the $\nu$-process the [F/O] ratio does not vary strongly with metallicity. Thus the differences observed in [F/O] cannot be explained by variations in the metallicity of the progenitors.
The neutrino luminosity may be small in the case of faint supernovae with a large black hole, which give high [$\alpha$/Fe], but there is no significant difference seen in the [$\alpha$/Fe] ratio between field halo stars and GC stars.
One possible scenario is as follows:
in GCs, the contribution from low-mass supernovae is smaller than in the field.
Since the star formation occurs in a baryon dominated cloud with very high density, the initial star burst can be very intense.
After the initial star burst, because of the small gravitational potential, outflow winds are generated immediately after the explosion of massive supernovae, which may remove the contribution from low-mass supernovae.
The small production of $\alpha$ elements from low mass supernovae means that the [F/O] ratio can reach values as large as $\sim 0$.
In contrast, massive supernovae produce more $\alpha$ elements which results in [F/O] ratios of $\sim -0.5$, consistent with the observational data.

The dotted line shows an example of such a GC model, where the timescale and duration of star formation is set to be $0.04$ Gyr and $0.02$ Gyr, respectively.
This model gives the MDF peaked at [Fe/H] $\sim -1.5$. The [(Mg,Si,S,Ca)/Fe] is as large as in the solar neighborhood model, which is consistent with observations of GCs in our Galaxy \citep[e.g.,][]{pri05}.
In a given GC, there is a spread in the observed O and F abundances, reflecting the so-called O-Na anti-correlation \citep[e.g.,][]{kra97,car09};
there is a primordial population with high O and N along with low Na and Al, and a polluted population with low O and N along with high Na and Al.
The polluted stars also include the products of H-burning at high temperature ($\sim 6.5 \times 10^7$ K), possibly from AGB stars or rotating massive stars \citep{gra04}.
Therefore, in Fig 1, we construct a model for GCs to fit the most O-rich stars, rather than the middle of the distribution.
For the stars in M22 (asterisks) the pollution from AGB stars seems to be large.

Note that the lack of low-mass supernovae is the opposite to the situation for dwarf spheroidal galaxies (dSphs) which have low [$\alpha$/Fe] and low [Mn/Fe] \citep{kob06}.
In dSphs, the dark matter component is large, the gas density is low, the star formation rate is low, and thus the contribution from massive supernovae is expected to be smaller than in the Milky Way halo.
We do not include the peculiar stars with s-process contribution \citep{abia10,abia11} and stars in the Large Magellanic Could \citep[e.g.,][]{cun03} because in the first case F is produced by AGB stars, and in the second case the chemical evolution in the LMC may be quite different from the Milky Way.

\section{Conclusions and Discussion}

Both the $\nu$-process of core-collapse supernovae and AGB stars play an important role in the production of fluorine.
We succeed in reproducing the observed F abundances with our chemical evolution model that includes the $\nu$-process of $E_\nu=3 \times 10^{53}$ erg.
At low metallicities ([O/H] $\ltsim -1.2$) F production is dominated by supernovae, and thus future observations of field stars at low-metallicities are important for constraining the neutrino luminosity released from a core-collapse supernova.
If the neutrino luminosity is specified, the F abundance along with C could be a good clock in the study of galactic archaeology to distinguish the contribution from AGB stars and supernovae.
The F observations of stars in GCs suggest that the star formation and chemical enrichment histories of GCs are different from those of field stars and that low-mass supernova played a smaller role in shaping the chemical evolution of these systems.

The $\nu$-process is also expected to be the producer of other elements such as K, Sc, and V.
With $E_\nu=9 \times 10^{53}$ erg, [(K,Sc,V)/Fe] ratios are increased to be closer to the observational data, but such a large improvement is not seen with the standard value of the neutrino luminosity.
There are several uncertainties that should be discussed;
for K, the NLTE correction in the observations is significant \citep{kob06}.
The Sc yields could also be increased by the low-density models that mimic 2D calculations \citep{ume05}.
There are also uncertainties in the reaction rates for V that may affect the nucleosynthesis calculations.

\acknowledgments
We would like to thank M. Lugaro and A. Alves-Brito for fruitful discussions.
This work was supported by the NCI National Facility at the Australian National University, the Global COE Program ``the Physical Sciences Frontier'', MEXT, Japan, Grants-in-Aid for JSPS Fellows(22-7342) and for Scientific Research (C) 23540287 of JSPS, and the Institute for the Physics and Mathematics of the Universe, University of Tokyo.

\end{document}